\documentclass[fleqn,twoside]{article}
\usepackage{espcrc2}

\usepackage{graphicx}

\newcommand{\ovl}[1]{\overline{#1}}

\newcommand{\p}{\partial}

\newcommand{\pslash}{p\kern-1ex /}
\newcommand{\lslash}{l\kern-1ex /}
\newcommand{\kslash}{k\kern-1ex /}
\newcommand{\dslash}{\p\kern-1.2ex /}
\newcommand{\Dslash}{{\cal D}\kern-1.5ex /}
\newcommand{\Aslash}{A\kern-1.2ex /}

\newcommand{\vev}[1]{\left\langle #1 \right\rangle}

\title{Calculation of weak matrix elements in domain-wall QCD with the
DBW2 gauge action}

\author{J.~Noaki\address[RBRC]{RIKEN-BNL Research Center, Brookhaven 
National Laboratory, Upton, NY 11973-5000, USA}\ \  for the RBC
Collaboration\thanks{Current members of the collaboration 
are: Y.~Aoki, T.~Blum, N.~Christ, M.~Creutz, C.~Dawson, T.~Izubuchi, 
L.~Levkova, X.~Liao, G.~Liu, R.~Mawhinney, Y.~Nemoto, J.~Noaki, S.~Ohta, 
K.~Orginos, S.~Prelovsek, S.~Sasaki and A.~Soni.}}
       
\begin{document}

\begin{abstract}
We report the details of our ongoing quenched calculations of weak 
matrix elements using the combination of domain-wall fermions and 
the DBW2 gauge action on lattices with $a^{-1}\approx 3$ GeV.
A strategy to avoid the problem of fixed topological charge 
is introduced in generating gauge configurations.
After studying the basic run parameters and elemental quantities,
we present a preliminary result for the kaon B-parameter ($B_K$).  
\vspace{1pc}
\end{abstract}

\maketitle

\section{Introduction}

The combination of domain-wall fermions and renormalization 
group (RG) improved gauge actions has been proven to yield fermions
with good chiral behavior, allowing progress in
the calculation of weak matrix
elements (WME) on the lattice ~\cite{Ishizuka}.
On the other hand, two examples of open issues are the scaling 
behavior of $B_K$ and the effect of the charm quark which contributes 
to $K\to\pi\pi$ matrix elements through the quark loop contraction
~\cite{Bob}. 
Both of these require the calculation of WME, using fermions with good
chiral behavior, on finer lattices than 
have been employed so far.

In this article, we present the current results from an ongoing quenched
numerical simulation by the RBC Collaboration.
It uses domain-wall fermions (DWF) and the DBW2 gauge action at a scale 
$a^{-1}\approx 3$ GeV with physical spatial size $\approx$ 1.6 fm.
Though the use of the RG-improved DBW2 gauge action improves
chiral symmetry, it has been recognized that the
topological charge of DBW2 gauge configurations generated by standard
Monte Carlo techniques evolves very slowly~\cite{Kostas}.
This fact motivated us to implement a strategy to generate gauge 
configurations whose topological charge is well-distributed \cite{foot}.

In the following, we discuss the details of our numerical simulation
and present preliminary results for the meson spectrum, residual chiral 
symmetry breaking mass $m_{\rm res}$ and $B_K$.

\section{Gauge configuration generation}

\begin{figure}[t]
\vspace{9pt}
\hspace{-0.1cm}\includegraphics[width=6.8cm, clip]{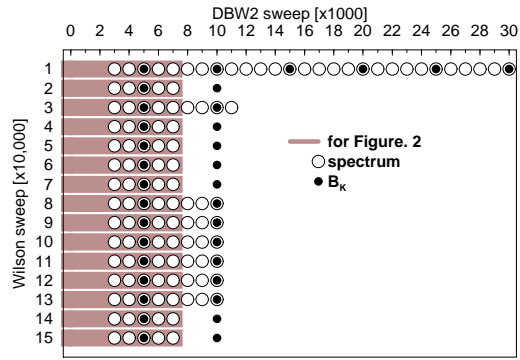}

\vspace{-0.5cm}

\caption{The gauge configurations used for spectrum (open circle)
 and $B_K$ (filled circle). Shadowed bars indicate data employed
in Figure~\ref{mp_var}.}
\label{status}
\end{figure}

On a $24^3\times 48$ lattice, we generate gauge configurations using
the Wilson action with $\beta =6.25$ in steps of 10,000 sweeps and use each 
of these as the initial configuration for
a DBW2 evolution with $\beta =1.22$ in steps of 1000 sweeps. 
The status of our numerical simulation for 15 independent DBW2 series is 
illustrated in Figure~\ref{status}.

The initial Wilson configurations have $a^{-1} \approx 3$ GeV, the same
lattice spacing desired for the DBW2 ensemble.  The part of the
topological charge produced by physical, long distance properties of
the Wilson lattices should be very close to that of a fully sampled
DBW2 ensemble, since at these weak couplings such a physical quantity
is independent of the UV details of the action.  We rely on this
physical argument to allow us to use the topological distribution
generated by the Wilson action as a starting point for our DBW2
ensemble.  With the 15 Wilson configurations discussed here, we find
$\vev{Q_{\rm top}}= -0.60\pm 3.11$. (For 50 configurations, we have
$\vev{Q_{\rm top}} =-0.32\pm 3.36$, with $-10\le Q_{\rm top}\le +7$.)
For a few of the DBW2 evolutions, $Q_{\rm top}$ changed by one unit,
while for the longest 30,000 sweep evolution, $Q_{\rm top}$ did not
change.  Thus, for the DBW2 action, we conclude that
over-relaxed and heat-bath steps would not generate a reasonable
sampling of topologies at these weak couplings.

\section{Hadron spectrum}
\newcommand{\cc}[1]{\multicolumn{2}{c}{#1}}
\begin{table}[t]
\caption{Simulation parameters for observables.}
\label{table1}

\begin{tabular}{c|c|c}
\hline
      & spectrum & $B_K$ \\
\hline\hline 
 size & \cc{$24^3\times 48$}\\
\hline
$M_5$ & 1.7 & 1.65\\ 
\hline
$L_s$ & 8 &  10 \\
\hline
$m_fa$ &  0.02, 0.03, 0.04 & 0.008 -- 0.040 \\
      &                   & in step of 0.008 \\
\hline
\#sweep & every 1000 & every 5000\\
\hline
\#config. & 120 & 30\\
\hline
\end{tabular}
\vspace{-0.4cm}

\end{table}

\begin{table}[b]
\caption{Spectrum results for $M_5=1.7$, $L_s=8$. }
\label{table2}
\begin{tabular}{cc}
\hline
$m_{PS}/m_V$ & 0.612(11), 0.698(9), 0.756(7) \\
$a^{-1}$ & 2.89(12) GeV \\
$m_sa/2$ & 0.0157(11) \\
$m_{\rm res}$ & 1.011(35) MeV \ ($M_5  = 1.7,L_s = 8$) \\
              & 0.276(10) MeV \ ($M_5 \! = \! 1.65,L_s \! = \! 10$) \\
\hline
\end{tabular}
\end{table}

\begin{figure}[t]
\vspace{9pt}
\includegraphics[width=6.9cm, clip]{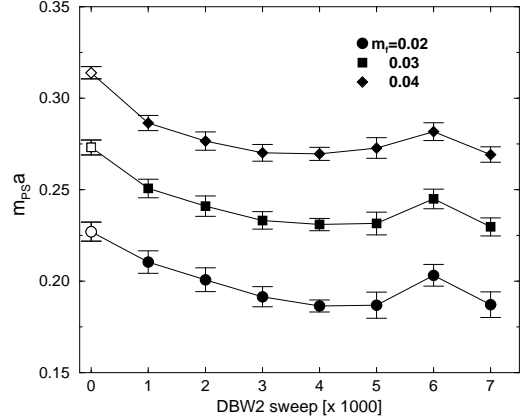}
\vspace{-0.8cm}

\caption{$m_{PS}a$ versus evolution in the DBW2 direction
 (filled symbols) obtained from averaging over 15 configurations in
 the vertical direction shown in
 Figure~\ref{status}. Open symbols represent the data from the initial
 Wilson configurations.}
\label{mp_var}
\vspace{-0.2cm}

\end{figure}

Simulation parameters for the spectrum calculation are 
summarized in the middle column of Table~\ref{table1}.  
To investigate how our strategy works and determine optimal parameters 
for the calculation of WME, we studied meson masses and the residual quark 
mass $m_{\rm res}$ which is induced by the explicit chiral symmetry
breaking due to finite $L_s$.

It is important to estimate the equilibration time for the DBW2
evolutions, starting from the initial Wilson configurations.
Figure~\ref{mp_var} shows the results for the pseudoscalar meson mass 
$m_{PS}a$ for each $m_fa$ and averaged over the configurations indicated 
in Figure~\ref{status}.
In this figure, one observes that about 3000 sweeps are necessary for
thermalization, which is consistent with a similar analysis for 
$m_{\rm res}$.  Our hadron spectrum determination only uses data from
3000 sweeps on.

Table~\ref{table2} contains the results from our spectrum calculation with 
$M_5=1.7$ and $L_s=8$.
The lattice scale $a^{-1}$ was determined from the rho meson
mass at the chiral limit.  We 
obtained the strange quark mass from the relation $m_{PS}^2=2B_0m_f$.
Further study of $m_{\rm res}$ as a function of domain-wall
height $M_5$, gave $1.65$ as the optimal value. 
Using this, we found $m_{\rm res}< 0.3$ MeV for the DBW2 action with 
$L_s = 10$.  This value for $L_s$ is used in the calculation of WME.

\section{Kaon B parameter}

\begin{figure}[t]
\includegraphics[width=6.8cm, clip]{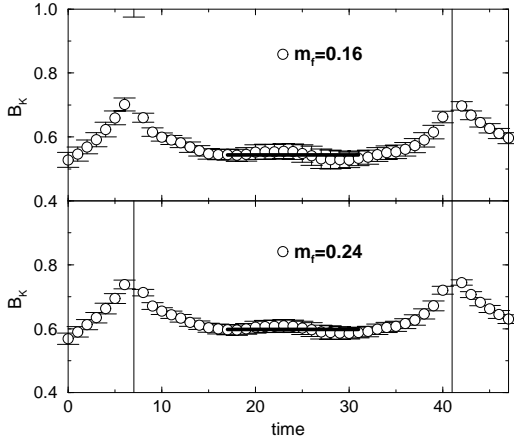}

\vspace{-0.9cm}

\caption{time dependence of $B_K$ for $m_fa= 0.016$ (upper) and $0.024$ 
 (lower). The solid lines indicate constant fits.}
\label{t-dep}
\end{figure}

\begin{figure}[t]
\vspace{-0.2cm}

\includegraphics[width=7.2cm, clip]{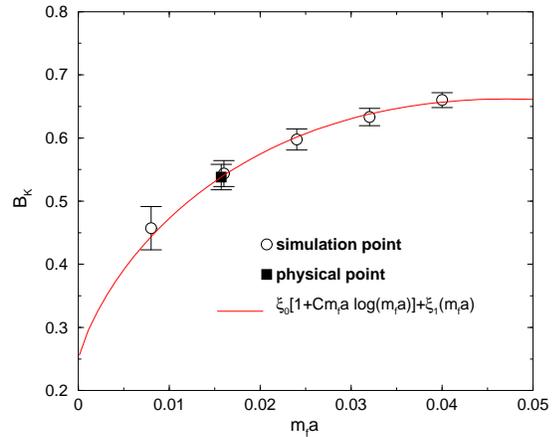}

\vspace{-0.9cm}

\caption{Lattice value of $B_K$ as a function of $m_fa$. The
 interpolated point (filled symbol) indicates the value at $m_fa = m_sa/2$. 
The solid line denotes a fit that is described in the text. }
\label{BK}
\end{figure}
\[
 B_K= \frac{\vev{\ovl{K}^0\Bigl|\bar{s}\gamma_\mu(1-\gamma_5)d\
\bar{s}\gamma_\mu(1-\gamma_5)d\Bigr|K^0}}
{\frac{8}{3}\vev{\ovl{K}^0\Bigl|\bar{s}\gamma_4\gamma_5d\Bigr|0}
\vev{0\Bigl|\bar{s}\gamma_4\gamma_5d\Bigr|K^0}}
\]
was calculated with the parameters summarized in the right column of 
Table~\ref{table1}.
Quark propagators from Coulomb gauge fixed spatial wall sources located at $t=$ 7
and 41 were used in our calculation and each propagator is an average 
of ones with periodic and anti-periodic boundary conditions in the time 
direction.

One finds reasonable plateaus in $t$ for $B_K$ as shown 
in Figure~\ref{t-dep}.
Figure~\ref{BK} shows the unrenormalized value of $B_K$ as a function 
of $m_fa$ and 
a fit to the function $B_K(m_fa)=\xi_0[1+Cm_fa\ln(m_fa)]+\xi_1m_fa$,  
where the value of $C$ is taken from chiral perturbation theory~\cite{Steve,GP}. 
The data are well fit to this function.
An uncorrelated fit has $\chi^2/{\rm dof}=0.14$.
The interpolated value $B_K= 0.539(21)$ at the physical point $m_fa=m_sa/2$ 
is indicated by a filled square. 
We have not yet calculated the non-perturbative renormalization factor 
for $B_K$. Employing the perturbative one~\cite{AIKT},
the value of $B_K$ in the $\ovl{\rm MS}$ NDR scheme can be estimated.

Previous works calculated $B_K(\mu=2\ {\rm GeV})$ with 
DWF in similar physical spatial volumes~\cite{CPPACS_BK,RBC_BK}. 
While our preliminary result is consistent with both of these within 
roughly one and two standard deviations, respectively, 
it is smaller than results from larger physical 
volume~\cite{CPPACS_BK}.

We thank RIKEN, BNL and the U.S.\ DOE for providing the facilities 
essential for the completion of this work.

\newcommand{\NP}{Nucl.~Phys.}
\newcommand{\NPSup}{Nucl.~Phys.~{\bf B} (Proc.~Suppl.)}
\newcommand{\PL}{Phys.~Lett.~}
\newcommand{\PR}{Phys.~Rev.~}
\newcommand{\PRL}{Phys.~Rev.~Lett.~}


\begin{thebibliography}{15}
\bibitem{foot}We changed a part of data presented in the talk to correct a
simple human error made in the random numbers used in generating the gauge configurations.
\end{thebibliography}
\end{document}